\documentstyle[epsfig]{elsart}

\newcommand{\bd}[1]{ \mbox{\boldmath $#1$}  }

\begin{document}

\begin{frontmatter}
\title{Dipole excited states in $^{11}$Li with complex scaling}

\author{E. Garrido} 
\address{Instituto de Estructura de la Materia, CSIC, Serrano 123, 
         E-28006 Madrid, Spain} 
\author{D.V.~Fedorov and A.S.~Jensen} 
\address{Institute of Physics and Astronomy, Aarhus
University, DK-8000 Aarhus C, Denmark}

\maketitle

\begin{abstract}
The 1$^-$ excitations of the three--body halo nucleus $^{11}$Li are
investigated. We use adiabatic hyperspherical expansion and solve the
Faddeev equations in coordinate space.  The method of complex scaling
is used to compute the resonance states.  The Pauli forbidden states
occupied by core neutrons are excluded by constructing corresponding
complex scaled phase equivalent two-body potentials.  We use a
recently derived neutron--core interaction consistent with known
structure and reaction properties of $^{10}$Li and $^{11}$Li.  The
computed dipole excited states with $J^\pi=1/2^+$, $J^\pi=3/2^+$, and
$J^\pi=5/2^+$ have energies ranging from 0.6 MeV to 1.0 MeV and widths
between 0.15 MeV and 0.65 MeV.  We investigate the dependence of the
complex energies of these states on the $^{10}$Li spectrum. The finite
spin 3/2 of the core and the resulting core-neutron spin-spin
interaction are important.  The connection with Coulomb dissociation
experiments is discussed and a need for better measurements is pointed
out.
\end{abstract}

\end{frontmatter}
\par\leavevmode\hbox {\it PACS:\ } 21.45.+v, 25.10.+s, 25.60.-t,  25.60.Gc

\section{Introduction and motivation}

During the last years a large effort has been made to investigate the
structure of two-neutron halo nuclei whose most prominent examples
are the Borromean systems $^{6}$He and $^{11}$Li. Their ground-state
properties are well reproduced by use of three--body models that
describe these nuclei as systems made of an inert core and
two--neutrons \cite{zhu93}. These models are also able to reproduce
many of the observables measured after fragmentation reactions
\cite{gar01}.

Nevertheless several uncertainties concerning especially the structure
of $^{11}$Li still remain. In a recent work \cite{gar02} the structure
of the two--body unbound system $^{10}$Li and its effect on the
overall $^{11}$Li structure has been investigated. Another question
still open refers to the continuum excited spectrum of $^6$He and
$^{11}$Li (these systems do not have bound excited states).  For this
investigation an appropriate treatment of the three--body wave
functions in the continuum and in particular of the resonant states is
needed.

Since the interactions involved in $^6$He are very well known this
nucleus is a perfect test case for the different models describing the
properties of two--neutron halo nuclei.  The $2^+$ resonance with
energy and width $(E_R,\Gamma_R)$=(0.822$\pm$25, 0.113$\pm$20) MeV
\cite{ajz89} is well established corresponding to an excitation energy
of around 1.8 MeV, as also confirmed in recent $^6$He fragmentation
reaction experiments \cite{aum99,ale00}.  Different theoretical
calculations reproduce the properties of this $2^+$ state
\cite{ers97,dan98}, although they differ in the predicted levels at
higher excitation energies.

Coulomb dissociation of $^{11}$Li on Pb at 28, 43 and 280 MeV/nucleon
shows a prominent peak at about 1 MeV in the differential
$B(E1)$-distribution in several experiments\cite{sac93,shi95,zin97}.
Unfortunately the results of these experiments differ significantly
from each other.  An excited state at roughly the same excitation
energy, $E^\star$=1.25$\pm$0.15 MeV \cite{kor96}, and consistent with
the quantum numbers $J^\pi$=0$^-$ or 1$^-$ \cite{kor97} was later
found by studying $^{11}$Li+$p$ collisions.  Together with this level
other states appeared at $\sim 3.0$, 4.9, 6.4, and 11.3 MeV.  A more
recent experiment \cite{gor98} investigating the pion capture reaction
$^{14}$C$(\pi^-,pd)^{11}$Li identified excited states for $^{11}$Li at
excitation energies 1.02$\pm$0.07, 2.07$\pm$0.12, and 3.63$\pm$0.13
MeV.

On the theoretical side only a rather small number of published
calculations of the $^{11}$Li spectrum exists. These continuum
three--body calculations are technically demanding not least due to
the inevitable effects of the hyperfine structure arising from the
non-zero spin of the $^{9}$Li core. In addition the neutron-core
interaction is uncertain since the spectrum is only incompletely known
and the parameters then must be determined indirectly by adjustments
to reproduce related observables. A full-scale calculation with
realistic parameters was reported in \cite{cob98}. A number of
$S$-wave poles were found with widths comparable to the real parts of 
their energies.  These poles arise
from large distances beyond 130~fm. They are not obtained in any other
published computation but on the other hand these large distances were
not accurately incorporated in any other numerical investigation.
The method in \cite{cob98} remains on the real radial coordinate axis
where the asymptotic boundary condition is enforced by matching the
exponentially increasing and oscillating resonance functions at large
distance to the proper combination of Hankel functions. This is
numerically very difficult and large distance properties must be
computed with high accuracy. Alternatives to test or to simplify the
computations would be very valuable.

A paper about the resonances of $^{11}$Li appeared recently
\cite{kum02}. The excited states were obtained in a three--body model
with separable potential by extension into the continuum.  The three
lowest excited states of $^{11}$Li given in \cite{gor98} are
reproduced without specification of angular momentum and parity as if
they were resonance energies measured from threshold.  In \cite{and99}
the effect of the dipole polarizability on the $B(E1)$ strength in
$^{11}$Li is investigated.  The E1 strength was also investigated in
\cite{bon97}, and more recently in \cite{for02}.  The agreement with
the experimental $B(E1)$--strength function is satisfactory, but the
excited $^{11}$Li spectrum is not computed.  These calculations were
performed with rather simplified $^{11}$Li wave functions where the
spin of the core is assumed to be zero. For $1^-$ excitations this is
an important deficiency since the hyperfine splitting can allow low
lying dipole excited states with one neutron in each of the lowest
neutron-core $s$ and $p$-states. These three states are very sensitive
to the spin splitting of the pairs of $(1^-,2^-)$ and
$(1^+,2^+)$-states in $^{10}$Li.  Models without this feature can then
be rather misleading in discussions of both the $^{10}$Li and
$^{11}$Li spectra.

One of the most efficient methods to calculate resonances, introduced
in the early 70's, is the complex rotation or complex scaling method
\cite{agu71,bal71,sim72}. The radial coordinates are rotated into the
complex plane by an angle $\theta$ ($r\rightarrow r
e^{i\theta}$). When this angle is larger than the argument of a
resonance then the divergent wave function corresponding to the
resonance becomes convergent, and its rotated wave function shows up
after solving the corresponding Schr\"{o}dinger equation in the same
way as a bound state. The details of this method and its application
to atomic collisions can be found in \cite{ho83,moi98}.

Application of the complex rotation method to investigate three--body
resonances has already been considered \cite{cso94,wan94}, and in
particular it has been used to investigate different three--body
nuclear systems. In \cite{cso96} the three--body resonances in $^3$H
and $^3$He have been investigated. The same method has also been
applied to study the continuum structure of $^9$Be and $^9$B
\cite{ara96} assuming these nuclei as three--body systems made by two
$\alpha$--particles and a neutron and a proton,
respectively. Application to the two--neutron halo nucleus $^6$He can
be found in \cite{cso94b,aoy95,myo01}, and in \cite{aoy97} the case of
$^{10}$He is considered. When available, the agreement with the
experimental data is satisfactory.  It is remarkable that, even if the
method has been applied during the last eight years to different
nuclear systems, there are no calculations available concerning the
halo nucleus $^{11}$Li, whose discovery is widely considered as the
starting point of nuclear halo investigations. Only one paper investigating 
the 3/2$^-$ excitations in $^{11}$Li with the complex scaling method has
very recently been published \cite{aoy02}. In this work the authors
point out that the experimental determination of the $p$-wave resonant
state and the $s$-wave virtual state in $^{10}$Li should make the
binding mechanism in $^{11}$Li clearer. They confirm the previously
predicted roughly equal amount of $s$ and $p$-wave content within the
$^{11}$Li \cite{gar99b}.  The 3/2$^-$ states investigated in
\cite{aoy02} correspond to $0^+$ excitations, meaning that splitting
of the two $p$--resonances and the two virtual $s$--states in
$^{10}$Li play a minor role. As pointed out above, a proper n-$^9$Li
interaction and an adequate treatment of the spin splitted states in
$^{10}$Li, are essential ingredients in descriptions of odd parity
excitations in $^{11}$Li as treated in the present paper.

A suitable method to describe spatially extended and weakly bound
three--body systems was introduced in \cite{fed93,fed94}. In this
method the Faddeev equations are solved in coordinate space by means
of the hyperspherical adiabatic expansion. This method is able to
reproduce accurately the asymptotics of the wave functions as proved
by the fact that it derives the Efimov effect in which an accurate
computation of the wave functions at very large distances is essential
\cite{fed93}. A detailed description of the method and applications to
atomic and nuclear physics has recently been published
\cite{nie01}. Halo nuclei, as weakly bound and extended systems, are
especially appropriate to be described by this method. During the last
years we have been investigating both the structure of the $^6$He and
$^{11}$Li ground states and the observables in fragmentation reactions
\cite{gar01,fed95,gar98}. The agreement with the experimental data is
satisfactory.

Combining the two successful methods of hyperspherical adiabatic
expansion and complex scaling is tempting, see \cite{fed02} for
details.  The recently established set of model parameters agreeing
with known properties of both $^{10}$Li and $^{11}$Li now allows to
fill the gap and compute the continuum spectrum of $^{11}$Li with all
the decisive features included.  The purpose of this paper is to
investigate the 1$^-$ excitations in $^{11}$Li, i.e.  the 1/2$^+$,
3/2$^+$, and 5/2$^+$ continuum states of $^{11}$Li obtained by 1$^-$
excitation from to the ground state of 3/2$^-$.

We start in section \ref{sec1} by describing very briefly the main
features of the complex scaling method. The specification to the
three--body case is shown in section \ref{sec1b}, where a brief
summary of the hyperspherical adiabatic method is given. The fact that
the core of the nucleus ($^9$Li) is a composite system containing
neutrons requires a careful treatment of the Pauli principle. We avoid
the neutrons from the halo to occupy Pauli forbidden levels by use of
the corresponding phase equivalent potentials, as explained in details
in section \ref{sec2}. In section \ref{sec3} we provide the
interactions used in the description of $^{11}$Li and the structure of
the ground state to which they give rise.  In section \ref{sec4} we
show the results for the 1$^-$ excitations of $^{11}$Li and we finish
in section \ref{sec5} with summary and conclusions.

\section{The complex scaling method} \label{sec1}

For a two--body system the partial wave expansion of the scattering
wave function is written as
\begin{equation}
\Psi(\bd{p}, \bd{r})=\sqrt{\frac{2}{\pi}} \sum_\ell i^\ell \frac{u_\ell(p,r)}{p r}
\sum_m Y_{\ell, m}(\Omega_r) Y_{\ell, m}^\star(\Omega_p) \;,
\end{equation}
where $\bd{p}$ and $\bd{r}$ are the relative momentum and the relative
coordinate, whose directions are defined by the angles $\Omega_p$ and
$\Omega_r$, respectively.  The quantum numbers $\ell$ and $m$ are the
relative angular momentum and its third component. For simplicity we
consider particles without spin.  The radial wave functions are
obtained by solving the Schr\"{o}dinger equation with the
corresponding two--body interaction $V(r)$.  For two non--interacting
particles the radial wave function $u_\ell(p,r)/pr$ is the spherical
Bessel function $j_\ell(pr)$, and the previous expansion becomes the
usual partial wave expansion of a plane wave.

The asymptotic behavior of $u_\ell(p,r)$ is given by
\begin{equation}
\frac{u_\ell(p,r)}{pr} \longrightarrow \frac{1}{2} \left( h^{(2)}_\ell(pr)+
                S_\ell(p) h^{(1)}_\ell(pr)\right) \;,
\label{asym1}
\end{equation} 
where $h^{(1)}_\ell(pr)$=$j_\ell(pr)$+$i\eta_\ell(pr)$ and
$h^{(2)}_\ell(pr)$=$j_\ell(pr)$-$i\eta_\ell(pr)$ are the spherical
Hankel functions, and $S_\ell(p)$ is the $S$--matrix. The large
distance behavior of $h_\ell^{(1)}$ and $h_\ell^{(2)}$ is given by
$\exp{(i(pr-\ell\pi/2))}/ipr$ and $\exp{(-i(pr-\ell\pi/2))}/ipr$,
respectively, leading to the following asymptotic behavior for
$u_\ell(p,r)$:
\begin{eqnarray}
u_\ell(p,r) \rightarrow 
\frac{1}{2i} \left(e^{-i(pr-\ell\pi/2)} +
                S_\ell(p) e^{i(pr-\ell\pi/2)}  \right) \; .
\label{asym2}
\end{eqnarray}  

When the $S$--matrix is analytically continued into the region of
complex values of $p$ the resonances appear as poles of the $S$-matrix
in the lower half--plane of $p$ away from the imaginary axis.
Therefore for a resonance the asymptotic behavior of the radial wave
function in eq.(\ref{asym1}) is given by
\begin{equation}
\hspace*{-5mm}
u_\ell(p,r) \longrightarrow \frac{pr}{2} h^{(1)}_\ell(pr) \longrightarrow
 \frac{1}{2i} e^{i(pr-\ell\pi/2)} =  
\frac{1}{2i} e^{|p|r \sin{(\theta_R)}} e^{i(|p|r \cos{(\theta_R)}-\ell\pi/2)}\;,
\label{asym3} 
\end{equation} 
where $|p|$ and $\theta_R$ are the value and the argument of the
complex momentum $p$ ($p=|p| e^{-i\theta_R}$).

Since $\sin{(\theta_R)}$ is positive the corresponding radial wave
function of a resonance diverges exponentially. The wave function is
therefore non square integrable. This fact makes it rather difficult
to establish numerically the resonant $p$--value. It is not simple to
know precisely the $p$--value for which the asymptotic behavior of the
radial wave function is given only by the exponentially increasing
function $h_\ell^{(1)}(pr)$ without any contribution from the
exponentially decreasing term $h_\ell^{(2)}(pr)$.

This problem is solved by means of the complex scaling method in which
the radial coordinate $r$ is transformed into $r e^{i\theta}$. In
other words, the radial coordinate in the hamiltonian $H$ is rotated
into the complex plane by the angle $\theta$.  The asymptotic behavior
of the radial solutions of the rotated hamiltonian $H^{(\theta)}$ is
now given by the radial rotation of eqs.(\ref{asym1}) and
(\ref{asym2}).  In particular for a resonance, according to
eq.(\ref{asym3}), the asymptotic behavior of the rotated resonant
radial wave function $u_\ell^{(\theta)}(p,r e^{i\theta})$ is given by
\begin{equation}
u_\ell^{(\theta)}(p,r e^{i\theta}) \longrightarrow  
\frac{1}{2i} e^{-|p|r \sin{(\theta-\theta_R)}} 
                      e^{i(|p|r \cos{(\theta-\theta_R)}-\ell\pi/2)}
\label{eq5}
\end{equation}

Therefore as soon as the scaling angle $\theta$ is larger than
$\theta_R$ the resonant radial wave function decreases exponentially,
and it can be computed following the same procedure as for a bound
state. Nevertheless the presence of $\sin{(\theta-\theta_R)}$ in the
exponent makes the value $r_{max}$ at which the asymptotics is reached
clearly larger than for a ``normal'' bound state, for which the
exponent is simply $-|p|r$.

It is important to realize that one of the main consequences of
complex scaling is that while the original hamiltonian $H$ is
hermitian the complex rotated hamiltonian $H^{(\theta)}$ is not
\cite{moi98,moi78}.  All the properties and theorems relative to the
non--rotated hamiltonian are valid for $H^{(\theta)}$ but using the
c--product instead of the ordinary scalar product. The c--product is
defined as

\begin{equation}
(f(r)|g(r))=\langle f^\star(r) | g(r) \rangle =\int f(r) g(r) d^3r 
\end{equation}
and, for instance, while the non rotated hamiltonian $H$ satisfies that
$\langle f|H|g\rangle = \langle g|H|f\rangle$, the complex scaled
hamiltonian $H^{(\theta)}$ satisfies $( f|H^{(\theta)}|g) = 
(g|H^{(\theta)}|f)$.

In particular, bound states of $H$ appear also as ordinary bound
states of the scaled hamiltonian $H^{(\theta)}$ \cite{ho83}. However,
while the radial wave function $\psi$ describing the bound state
before rotation is normalized such that $\langle \psi|\psi\rangle=1$,
after rotation the bound state wave function $\psi^{(\theta)}$ is
normalized such that
\begin{equation}
\hspace*{-1cm}
( \psi^{(\theta)}(z)|\psi^{(\theta)}(z) ) = 
\int \psi^{(\theta)}(z)\psi^{(\theta)}(z) dz =  
e^{i\theta} \int_0^\infty \psi^{(\theta)}(re^{i\theta})
\psi^{(\theta)}(re^{i\theta}) dr =1 
\label{norm}
\end{equation}

\section{The three--body case} \label{sec1b}

To describe a three--body system we start by introducing the usual
Jacobi coordinates ($\bd{x}_i$, $\bd{y}_i$)
\begin{equation}
\hspace*{-0.8cm}
\bd{x}_i=\sqrt{\frac{1}{m} \frac{m_j m_k}{m_j+m_k}} (\bd{r}_i-\bd{r}_k) \;,
\bd{y}_i=\sqrt{\frac{1}{m} \frac{m_i(m_j+m_k)}{m_i+m_j+m_k}} 
                    (\bd{r}_i-\frac{m_j \bd{r}_j+m_k \bd{r}_k}{m_j+m_k}) \; ,
\end{equation}
where $\{i,j,k\}$ is a cyclic permutation of $\{1,2,3\}$, $m_i$,
$m_j$, and $m_k$ are the masses of the three particles in the system,
and $m$ is an arbitrary normalization mass that we take to be the
nucleon mass.

For each of the three possible sets of Jacobi coordinates the
hyperspherical coordinates are defined by the hyperradius
$\rho=\sqrt{x_i^2+y_i^2}$ and the five hyperangles
$\alpha_i=\arctan{(x_i/y_i)}$ and $\Omega_{x_i}$ and $\Omega_{y_i}$
that give the directions of $\bd{x}_i$ and $\bd{y}_i$.

The three--body wave function is computed by solving the Faddeev
equations following the hyperspherical adiabatic expansion described
in \cite{nie01}.  The wave function $\Psi$ is expanded in a complete
set of generalized angular functions $\Phi_n^{(i)}(\rho,\Omega_i)$
\begin{equation}
\Psi= \frac{1}{\rho^{5/2}} \sum_n f_n(\rho) 
\sum_{i=1}^3 \Phi_n^{(i)}(\rho,\Omega_i) \;,
\label{eq7}
\end{equation}
where the radial expansion coefficients, $f_n(\rho)$, are independent
of $i$ and the angular functions are the eigenvectors of the angular
part of the Faddeev equations:
\begin{equation}
\hspace*{-1cm}
\hat{\Lambda} \Phi_n^{(i)} + \frac{2m\rho^2}{\hbar^2}
V_{jk}(\rho \sin{\alpha_i}) (\Phi_n^{(i)}+\Phi_n^{(j)}+\Phi_n^{(k)})=
\lambda_n(\rho) \Phi_n^{(i)} \; , \; i=1,2,3
\end{equation}
where the angular operator $\hat{\Lambda}$ can be found for instance
in \cite{nie01} and $V_{jk}$ is the interaction between particles $j$
and $k$.

The radial coefficients $f_n(\rho)$ are obtained from a coupled set of
differential equations in which the angular eigenvalues
$\lambda_n(\rho)$ enter as effective potentials:

\begin{small}
\begin{equation}
\hspace*{-1cm}
\left[ \frac{d^2}{d\rho^2} - \frac{2mE}{\hbar^2} + \frac{1}{\rho^2} 
\left( \lambda_n(\rho)+\frac{15}{4} \right)\right] f_n(\rho)+ 
\sum_{n^\prime} \left(-2P_{n n^\prime}(\rho)
\frac{d}{d\rho}-Q_{n n^\prime}(\rho)   \right) f_{n^\prime}(\rho)=0 \; ,
\label{radial}
\end{equation}
\end{small}
where the functions $P$ and $Q$ also can be found in \cite{nie01}.

\subsection{Complex coordinate rotation}

For a system of three particles in the continuum the $S$--matrix has
known analytical properties similar to a two--particle multichannel
problem. The radial wave functions must then be labeled with two
indexes, one of them ($n$) being the usual index on the components,
and a second index $n^\prime$ labeling all the possible asymptotic
behaviors and therefore labeling different total wave functions in
eq.(\ref{eq7}). We can actually interpret $n$ and $n^\prime$ as labels
for the ingoing and outgoing three--body channels.

The asymptotic behavior of the radial three--body continuum wave 
functions is given by:
\begin{equation}
f_{nn^\prime}(\rho) \longrightarrow  
\sqrt{\frac{m\rho}{4\hbar^2}}\left(
   \delta_{nn^\prime} H^{(2)}_{K+2}(\kappa \rho) + 
                    S_{nn^\prime} H^{(1)}_{K+2}(\kappa \rho)
                                  \right) ,
\label{asym3b} 
\end{equation}
where $\kappa=\sqrt{2mE/\hbar^2}$, $E$ is the three--body energy and
$H^{(1,2)}_\mu(\kappa \rho)$ are the Hankel functions. 
The value of the hypermomentum $K$ is defined by the
asymptotic value of the eigenfunction $\lambda_n(\rho)$ associated to
channel $n$, that is given by $K(K+4)$.

Again, when the $S$--matrix is analytically continued into the complex
$\kappa$--plane resonances show up as poles of the $S$--matrix in the
lower half--plane and away from the imaginary axis. Therefore the
asymptotics of the radial resonance wave function is given only by the
Hankel function $H^{(1)}_{K+2}(\kappa \rho)$ that goes at large
distances as $\sqrt{2/\pi \kappa \rho}\exp{(i(\kappa \rho - K \pi/2
+3\pi/4))}$.  Writing now the complex wave number $\kappa$ as
$|\kappa| e^{-i\theta_R}$ we find for the asymptotics of
$f_{nn^\prime}(\rho)$ the expression
\begin{equation}
f_{nn^\prime}(\rho) \longrightarrow
e^{|\kappa|\rho \sin{(\theta_R)}} e^{i(|\kappa|\rho 
\cos{(\theta_R)}- K \pi/2 + 3\pi/4)} \; ,
\end{equation}
that is equivalent to eq.(\ref{asym3}).

Again the exponential divergence in the resonance radial wave function
can be eliminated by use of the complex scaling method. Rotation of
the Jacobi coordinates by an angle $\theta$ amounts to rotation of the
hyperradius $\rho$ into $\rho e^{i\theta}$, since all the five
hyperangles remain unchanged. Therefore, as in the two--body case, the
rotated radial wave function of the three--body resonance has the
asymptotic behavior given by
\begin{equation}
f_{nn^\prime}(\rho e^{i \theta}) \rightarrow  
e^{-|\kappa|\rho \sin{(\theta-\theta_R)}} 
  e^{i(|\kappa|\rho \cos{(\theta-\theta_R)}- K \pi/2 + 3\pi/4)} \;, 
\label{asym3br} 
\end{equation}
that decreases exponentially for values of $\theta$ larger than
$\theta_R$. As a consequence, combining the complex scaling with the
hyperspherical adiabatic expansion method described in \cite{nie01}
will permit calculation of the complex energy of a three--body
resonance ($E=E_R-i\Gamma_R/2$, where $E_R$ is the resonance energy
and $\Gamma_R$ is the width), as well as its rotated wave function.

After complex scaling the effective potentials $\lambda_n(\rho)$
entering in the radial part of the Faddeev equations
(eq.(\ref{radial})) become complex quantities, and depend as well on
the complex rotation angle $\theta$. As shown in \cite{nie01} at short
distances the non--rotated $\lambda$'s behave as
\begin{equation}
\lambda(\rho) \stackrel{\rho \rightarrow 0}{\longrightarrow} 
K(K+4)-b \rho^2 \;,
\label{lam0}
\end{equation}
where $K$ is the hypermomentum and $b$ is a positive constant for
attractive potentials. After transformation of $\rho$ into $\rho
e^{i\theta}$ we see that the real part of the rotated $\lambda$'s goes
as
\begin{equation} 
{\rm Re}(\lambda(\rho e^{i\theta})) \stackrel{\rho \rightarrow 0}
{\longrightarrow} K(K+4)-b \rho^2 \cos{(2\theta)} \;
\label{reall} 
\end{equation}
and therefore at $\rho=0$ the real part of the rotated $\lambda$-spectrum
corresponds to the hyperspherical spectrum $K(K+4)$, where for $1^-$
excitations $K$ has to be odd. In the same way the imaginary part of
the rotated $\lambda$'s goes like
\begin{equation}
{\rm Im}(\lambda(\rho e^{i\theta})) \stackrel{\rho \rightarrow 0}
{\longrightarrow} -b \rho^2\sin{(2\theta)}
\label{imal}
\end{equation} 
and therefore all of them start at zero and take negative values at
short distances.  We also see that the larger the rotation angle
$\theta$ the higher the slope of the imaginary part of the $\lambda$
function at short distances.

At large distances, assuming that we are dealing with Borromean
systems, the $\lambda$ functions behave as \cite{nie01}
\begin{equation}
\lambda(\rho) \stackrel{\rho \rightarrow \infty}{\longrightarrow}
K(K+4)-\frac{c}{\rho^{1+2\ell_0}} \;,
\label{laminf}
\end{equation}
where $c$ is a positive constant for attractive potentials and
$\ell_0$ is an integer.  We then see that after complex scaling the
real part of the rotated $\lambda$'s is again recovering the
hyperspherical spectrum at infinity, while the imaginary part goes
like
\begin{equation}
{\rm Im}(\lambda(\rho e^{i\theta})) \stackrel{\rho \rightarrow 
\infty}{\longrightarrow} c \sin{((1+2\ell_0)\theta)}/\rho^{1+2\ell_0} \; .
\label{iminf}
\end{equation}
Therefore the imaginary part of the complex rotated $\lambda$'s goes
to zero at infinity from above. Since at short distances the imaginary
part is negative we conclude that for all the rotated $\lambda$'s in the
expansion (\ref{radial}) the imaginary part has an oscillatory
behavior, starting at zero, becoming negative, crossing afterwards the
zero axis, and going finally again to zero at infinity from above.

\subsection{The Pauli principle and complex scaled phase equivalent 
potentials}
\label{sec2}

An accurate treatment of the Pauli principle is one of the most
important points when describing few--cluster systems where the
clusters are composite structures containing identical fermions.  In
\cite{gar99} we have shown that the use of phase equivalent potentials
\cite{suk85,bay87,fie90,bay87b} is an appropriate method to implement
the Pauli principle in three--body calculations based on the adiabatic
hyperspherical approach.

Phase equivalent potentials are constructed as follows.  Let us
consider two particles interacting with each other via a potential
$V(r)$ such that the two--body system has one or several bound states,
where the radial wave function of its lowest bound state is
$\psi_{\ell s j}(r)$. It is then possible to construct a second
potential with exactly the same phase shifts as $V(r)$ for any value
of the two--body energy and such that the new potential has the same
bound state spectrum except the one described by $\psi_{\ell s
j}(r)$, that is removed. This new phase equivalent potential has the form
\cite{bay87b}:
\begin{equation}
V^{(pe)}(r)=V(r) - \frac{(\hbar c)^2}{\mu} \frac{d^2}{dr^2} \ln
\left( \int_0^r |\psi_{\ell s j}(r^\prime)|^2 dr^\prime\right) \; ,
\label{pep}
\end{equation}
where $\mu$ is the reduced mass of the two particles.

Simultaneous application of the procedure described in \cite{gar99} to
treat the Pauli principle and the complex scaling method obviously
requires the complex scaling of the corresponding phase equivalent
potentials. After complex scaling the bound states of the original
potential are also present, and in particular those bound states
forbidden by the Pauli principle should be removed from the
calculation. To do this we construct the corresponding phase
equivalent potential where the forbidden state has been removed
according to eq.(\ref{pep}), but where the radial coordinate $r$ is
replaced by the scaled coordinate $z=re^{i\theta}$. Therefore the
complex rotated phase equivalent potential is given by:

\begin{eqnarray}
V^{(pe)}(z)&=&V(z) - \frac{(\hbar c)^2}{\mu} \frac{d^2}{dz^2} \ln
\left( \int_{z^\prime=0}^{z^\prime=z} \left(
\psi_{\ell s j}^{(\theta)}(z^\prime)\right)^2 dz^\prime \right) =
  \nonumber \\ && \hspace*{-1.5cm}
V(z) - \frac{(\hbar c)^2}{\mu} e^{-i\theta}\frac{d^2}{dr^2} \ln
\left( \int_{r^\prime=0}^{r^\prime=r} \left(
\psi_{\ell s j}^{(\theta)}(r^\prime e^{i\theta})\right)^2 dr^\prime\right)
\end{eqnarray}
where $\psi_{\ell s j}^{(\theta)}(r e^{i\theta})$ is the complex
rotated radial wave function of the Pauli forbidden bound state, that
is normalized according to eq.(\ref{norm}). After performing the
second derivative one gets the following final expression for the
complex scaled phase equivalent potential:
\begin{small}
\begin{equation}
\hspace*{-9mm}
V^{(pe)}(re^{i\theta})=  V(re^{i\theta}) -
\frac{(\hbar c)^2}{\mu} e^{-i\theta}    
\left[
\frac{2\psi_{\ell s j}^{(\theta)}(re^{i\theta})  
\frac{d\psi_{\ell s j}^{(\theta)}(re^{i\theta})}{dr} }
{ \int_{0}^{r} (
\psi_{\ell s j}^{(\theta)}(r^\prime e^{i\theta}))^2 dr^\prime }
- \frac{(\psi_{\ell s j}^{(\theta)}(re^{i\theta}))^4}
{ \left(\int_{0}^{r} (
\psi_{\ell s j}^{(\theta)}(r^\prime e^{i\theta}))^2 dr^\prime \right)^2 }
\right] \; .
\end{equation}
\end{small}

\section{Bound $^{11}$Li structure}
\label{sec3}

\begin{figure}
\begin{center}
\epsfxsize = 8cm
\epsfbox{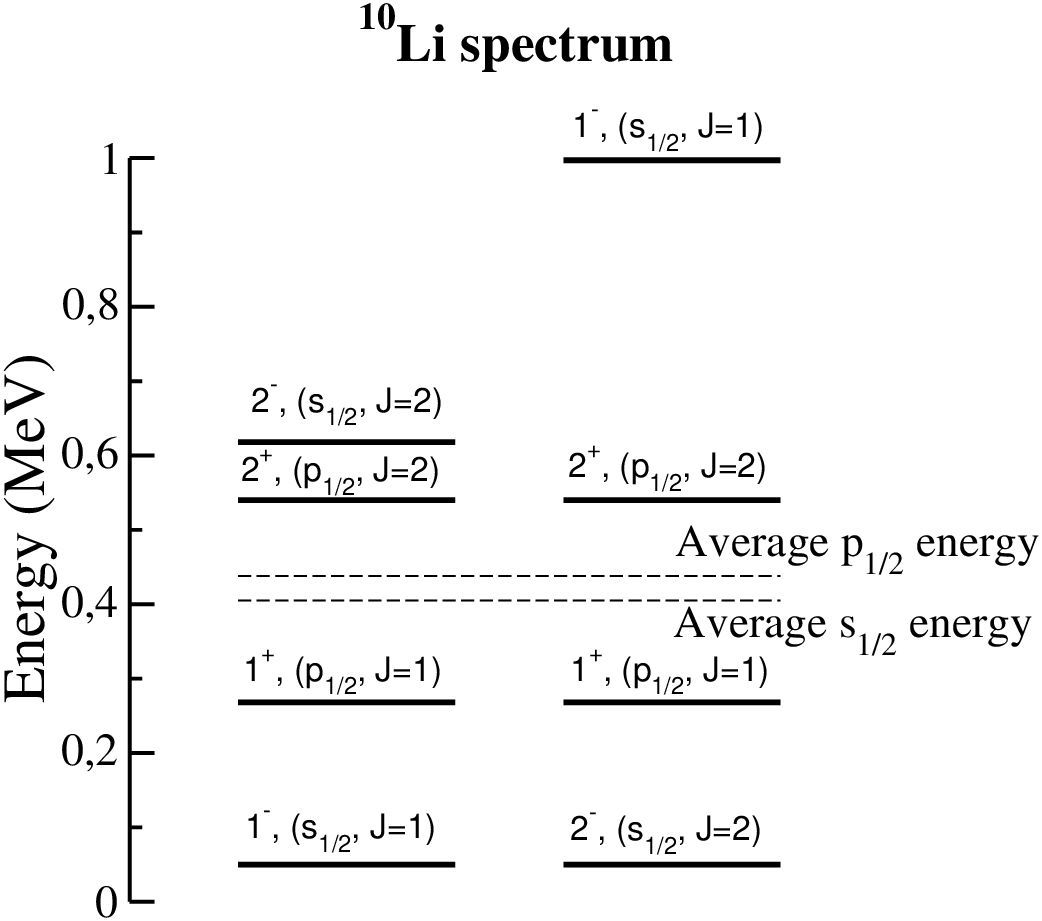}
\end{center}
\caption[]{$^{10}$Li spectra obtained with the neutron-$^9$Li interaction
used in the three--body calculation. }
\label{fig1}
\end{figure}

In ref.\cite{gar02} we have investigated the structure of the unbound
nucleus $^{10}$Li and its connection with the ground state properties
of $^{11}$Li. We have used a neutron--$^9$Li potential consistent with
the structure of the $^9$Li core and the energies and quantum numbers
of the experimentally established two--body resonances and virtual
states in $^{10}$Li. In particular the neutron--core interaction is
reproducing a 2$^+$ state at 0.54 MeV and a virtual state (either
1$^-$ or $2^-$ state) at 50 keV\footnote{
Energies of virtual s-states are strictly negative. However,
to avoid confusion with bound states we use the absolute values.
Therefore in the spectrum the virtual states then appear among
resonances although they do not show up in the cross section with
the characteristic resonance bump.
}. In this work we have then
investigated the range of energies of the 1$^+$ $p$--resonance and the
second virtual state still consistent with the known properties of
$^{11}$Li, i.e. the existence of one and only one bound state with
$J^\pi=\frac{3}{2}^-$, a binding energy of $\sim$0.300 keV, and a
$p$--wave content of $\sim$40\%.  In this way we have determined a set
of possible neutron--$^9$Li interactions for which we have computed
the $^{10}$Li invariant mass spectrum after fragmentation of $^{11}$Li
on a carbon target.  This observable is especially interesting because
it is probably the most sensitive one to the energy levels of each of
the two--body subsystems contained in the three--body
nucleus. Comparison of the computed and experimental invariant mass
spectra was then used to establish the most likely spectra for
$^{10}$Li. In ref.\cite{gar02} we have then concluded that together
with the 2$^+$ $p$--resonance at 0.54 MeV $^{10}$Li must have a second
$p$--resonance (1$^+$) at an energy of 0.35$\pm$0.15 MeV. For the
virtual states one of the two possible levels (1$^-$ or 2$^-$) is
placed at a low energy ($\sim$ 50 keV), while the second one must be
at an energy of 0.80$\pm$0.30 MeV.

To be precise in the present work we have used the potential
parameters corresponding to the spectra $c$ and $d$ in the upper part
of fig.6 in \cite{gar02}. These two spectra are shown in
fig.\ref{fig1}, and they are such that both of them have a
$2^+$--resonance at 0.54 MeV and a $1^+$--resonance at 0.25 MeV,
consistent with the $1^+$ level at 0.24$\pm$0.06 MeV observed in
\cite{boh97}. The difference between both spectra is that in one of
them the virtual $s$--state at 50 keV is the 1$^-$-- state (left part
of fig.\ref{fig1}), while for the other one it is the $2^-$--state.
Both spectra are consistent with all experimental information and only
small modifications are allowed to maintain this agreement.

These two spectra are obtained with a neutron-$^9$Li interaction of
the form
\begin{equation}
V_{nc}^{(\ell)}(r)=V_c^{(\ell)}(r)
   +V^{(\ell)}_{ss}(r) \langle \bd{s_n} \cdot \bd{s}_c \rangle
   +V^{(\ell)}_{so}(r) \bd{\ell}_{nc} \cdot \bd{s}_{n} \; ,
\end{equation}
where $\langle \bd{s_n} \cdot \bd{s}_c \rangle= \langle \ell_{nc},
j_{n}, J |\bd{s_n} \cdot \bd{s}_c| \ell_{nc}, j_{n}, J\rangle$,
$\bd{s}_n$ and $\bd{s}_c$ are the intrinsic spins of the neutron and
the core, $\bd{\ell}_{nc}$ is their relative orbital angular momentum,
$j_n$ is the coupled momentum of $\ell_{nc}$ and $s_n$, and $J$ is the
total angular momentum obtained after coupling of $j_n$ and the spin
of the core $s_c$.

In the calculation $s$--waves and $p$--waves have been included. The
radial central, spin--spin, and spin--orbit potentials are taken to be
gaussians with range equal to 2 fm. The strengths of the gaussians are
-94.0 MeV and -79.64 MeV for the central $s$ and $p$--potentials
respectively. For the spin--orbit and spin--spin $p$--potentials the
strength of the gaussians is -13.12 MeV and 1.10 MeV, respectively.
For the spin--spin potential for $s$--waves we take 6.85 MeV and
$-11.4$ MeV when the 1$^-$ and $2^-$ virtual states respectively are
placed at 50 keV.  The neutron--neutron interaction is given in
\cite{gar97}.

The potentials specified above are such that the
$s_{1/2}$--interaction has a low--lying virtual state at 50 keV and a
deeply bound state while the $p_{3/2}$--interaction has a bound state
at $-4.1$ MeV, that is the neutron separation energy in $^9$Li
\cite{ajz88}.  Since the $s_{1/2}$--shell and the neutron
$p_{3/2}$--shell are completely filled by the neutrons in the $^9$Li
nucleus these states are forbidden by the Pauli principle when adding
more neutrons as for $^{10}$Li and $^{11}$Li. As mentioned in the
previous section the Pauli principle is taken into account by
substituting the $s_{1/2}$ and the $p_{3/2}$ interactions by the
corresponding phase equivalent potentials.

Another feature to be considered is that the two--body interactions
alone underbind the three--body nucleus $^{11}$Li. This is a well
known general problem for few--body systems \cite{car98}. To recover
the experimental value of the binding energy we introduce a
phenomenological three--body interaction that accounts for the
polarization of the particles. The shape of this three--body force is
a gaussian in hyperradius with a range equal to 3 fm and a strength
-3.9 MeV.

\begin{table}
\caption{Components used in the calculation of the $J^{\pi} =
\frac{1}{2}^+$ state.  $K_{max}$ is the maximum value of the
hypermomentum $K$ used in the hyperspherical expansion. The left part
of the table refers to the components in the Jacobi set where $\bd{x}$
connects the two halo neutrons, while in the right part $\bd{x}$
connects the $^9$Li core and one neutron.  }
\begin{tabular}{c|ccc|cccccc}
\hline
$\ell_x$ & 0  &  1  &  1 & 0 & 0 & 0 & 1 & 1 & 1  \\
$\ell_y$ & 1  &  0  &  0 & 1 & 1 & 1 & 0 & 0 & 0  \\
     $L$ & 1  &  1  &  1 & 1 & 1 & 1 & 1 & 1 & 1  \\
   $s_x$ & 0  &  1  &  1 & 1 & 1 & 2 & 1 & 1 & 2  \\
     $S$ &3/2 & 1/2 & 3/2&1/2&3/2&3/2&1/2&3/2&3/2 \\
$K_{max}$&181 & 121 & 121& 121&181&121 & 121&121 &181  \\
\hline
\end{tabular}
\label{tab1}
\end{table}

\begin{table}
\caption{As in table \ref{tab1} for the $J^{\pi} = \frac{3}{2}^+$ state.}
\begin{tabular}{c|cccc|cccccccc}
\hline
$\ell_x$ & 0  &  1  &  1 & 1 & 0 & 0 & 0 & 0 & 1 & 1 & 1 & 1  \\
$\ell_y$ & 1  &  0  &  0 & 0 & 1 & 1 & 1 & 1 & 0 & 0 & 0 & 0  \\
     $L$ & 1  &  1  &  1 & 1 & 1 & 1 & 1 & 1 & 1 & 1 & 1 & 1  \\
   $s_x$ & 0  &  1  &  1 & 1 & 1 & 1 & 2 & 2 & 1 & 1 & 2 & 2  \\
     $S$ &3/2 & 1/2 & 3/2&5/2&1/2&3/2&3/2&5/2&1/2&3/2&3/2&5/2 \\
$K_{max}$&201 & 121 &121 &121&181&181&181&181&121&121&121&121 \\
\hline
\end{tabular}
\label{tab2}
\end{table}

\begin{table}
\caption{As in table \ref{tab1} for the $J^{\pi} = \frac{5}{2}^+$ state.}
\begin{tabular}{c|ccc|cccccc}
\hline
$\ell_x$ & 0  &  1  &  1 & 0 & 0 & 0 & 1 & 1 & 1  \\
$\ell_y$ & 1  &  0  &  0 & 1 & 1 & 1 & 0 & 0 & 0  \\
     $L$ & 1  &  1  &  1 & 1 & 1 & 1 & 1 & 1 & 1  \\
   $s_x$ & 0  &  1  &  1 & 1 & 2 & 2 & 1 & 2 & 2  \\
     $S$ &3/2 & 3/2 & 5/2&3/2&3/2&5/2&3/2&3/2&5/2 \\
$K_{max}$&181 & 121 & 121&181&181&181&121&121 &121  \\
\hline
\end{tabular}
\label{tab3}
\end{table}

\section{Dipole excited states}
\label{sec4}

To compute the 1$^-$ excited states in $^{11}$Li we follow exactly the
same procedure as for the $^{11}$Li ground state but with the complex
scaling transformation described in section \ref{sec1b}. We use then
the hyperspherical adiabatic expansion method to solve the complex
scaled Faddeev equations in coordinate space. The Faddeev equations
are solved with the complex scaled neutron--neutron and neutron--core
interactions described in the previous section. The Pauli principle is
taken into account as described in section \ref{sec2} by use of the
corresponding complex scaled phase equivalent potentials.

The $^{11}$Li ground state has the quantum numbers
$J^\pi=\frac{3}{2}^-$, and therefore a $1^-$ excitation will lead to
states with spin and parity $\frac{1}{2}^+$, $\frac{3}{2}^+$, or
$\frac{5}{2}^+$. The only difference between the calculations of these
three states is in the components included, since for each of them the
quantum numbers of each component have to be consistent with the total
spin and parity.  In tables~\ref{tab1}, \ref{tab2}, and \ref{tab3} we
specify the components used for the $\frac{1}{2}^+$, $\frac{3}{2}^+$,
and the $\frac{5}{2}^+$ states, respectively.  In each table the left
part shows the components used in the Jacobi set where the
$\bd{x}$--coordinate connects the neutrons, while in the right part
the components refer to the two Jacobi sets in which $\bd{x}$ connects
one of the neutrons and the core.

\subsection{Complex rotated hyperspherical adiabatic potentials}

\begin{figure}
\begin{center}
\epsfysize = 7cm
\epsfbox{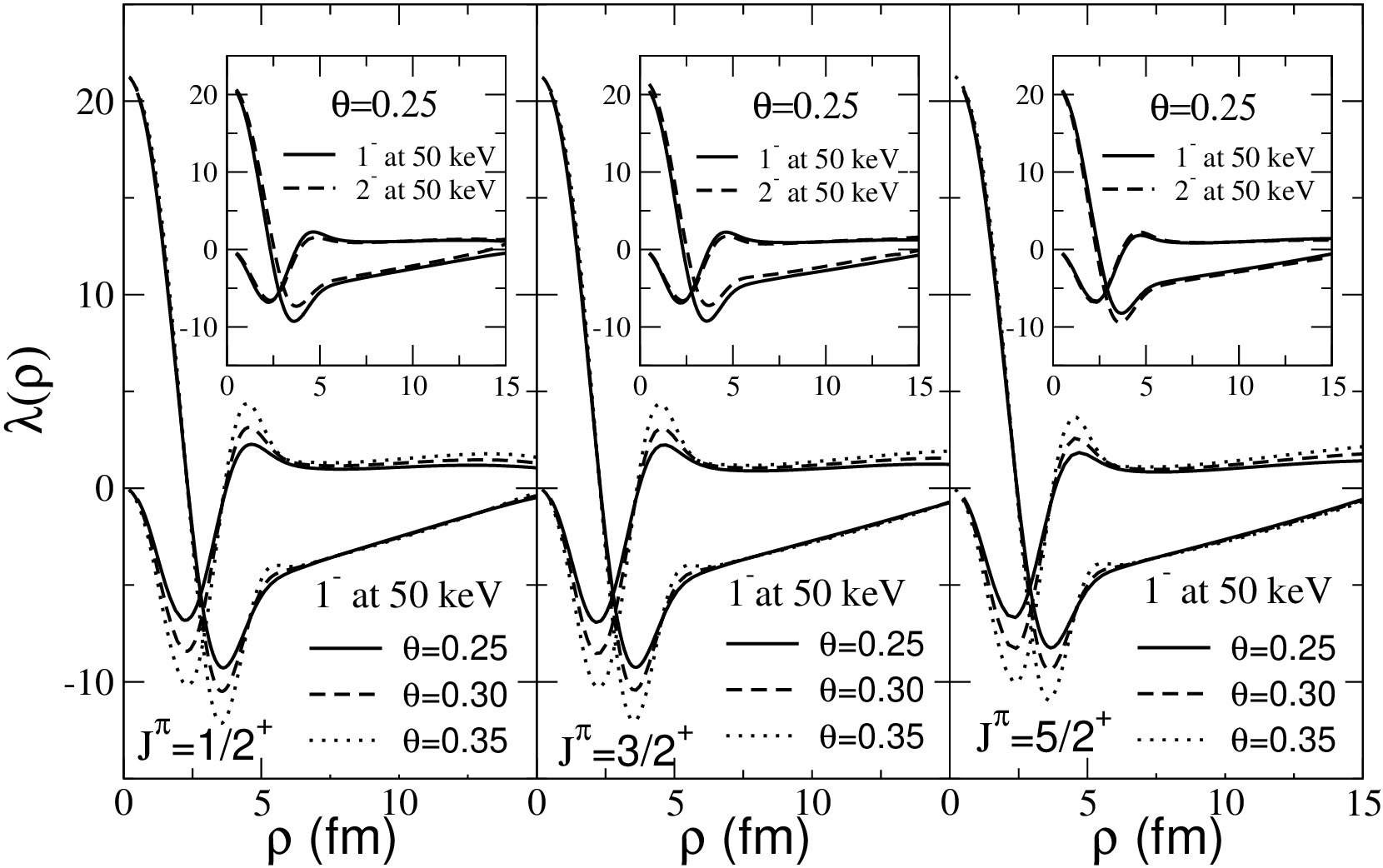}
\end{center}
\caption[]{Deepest rotated $\lambda$--function for the $1/2^+$ state (left)
the $3/2^+$ state (middle), and $5/2^+$ state (right) in $^{11}$Li.
In the external plots we show the complex $\lambda$'s for rotation
angles $\theta=0.25$ (solid), $\theta=0.30$ (dashed), and $\theta=0.35$
(dotted) for the $^{10}$Li spectrum in the left part of 
fig.\protect\ref{fig1}.
The real parts are the curves starting at 21, while the imaginary parts
start at zero. In the insets we compare for $\theta=0.25$ the deepest
$\lambda$ for the left (solid) and right (dashed) $^{10}$Li spectra in 
fig.\protect\ref{fig1}.  }
\label{fig2}
\end{figure}

The general behavior at short and large distances of the
$\lambda$--effective potentials entering in the radial equations
(\ref{radial}) is given by eqs.(\ref{lam0}) to (\ref{iminf}). This
general behavior can be observed in the complex scaled $\lambda$'s
obtained for the $1/2^+$, $3/2^+$, and $5/2^+$ states in $^{11}$Li. In
all the calculations shown in this work we have considered two
$\lambda$'s in the expansion (\ref{radial}), although inclusion of the
second $\lambda$ is not changing the results significantly. Typically
the second $\lambda$ is giving from 5 to 10\% of the rotated wave
function. In fig.\ref{fig2} we show the real and imaginary parts of
the most contributing $\lambda(\rho)$ for three different values of
the rotation angle $\theta=0.25$, $\theta=0.30$, and
$\theta=0.35$. The left, central, and right parts of the figure
correspond to the $1/2^+$ state, the $3/2^+$ state, and the $5/2^+$
state, respectively. The potentials used in the calculations produce
the $^{10}$Li spectrum described in section \ref{sec3}, i.e., a 1$^+$
$p$--resonance at 0.25 MeV, a 2$^+$ $p$--resonance at 0.54 MeV, and a
low--lying virtual $s$--state at 50 keV. The $\lambda$--functions
shown in the external parts of the figures correspond to the $1^-$
level in $^{10}$Li at 50 keV ($^{10}$Li spectrum in the left par of
fig.\ref{fig1}). In the inner part of the figures we compare for
$\theta=0.25$ the lowest $\lambda$ functions when the 1$^-$ (solid
line) and the $2^-$ (dashed line) is at 50 keV.

According to eqs.(\ref{reall}) and (\ref{laminf}) the real part of the
rotated $\lambda$'s have to coincide with the hyperspherical spectrum
$K(K+4)$ at both $\rho=0$ and $\rho=\infty$.  Since we are considering
negative parity states only odd values of $K$ are then possible. In the
figure the real part of the $\lambda$'s is given by the curves
starting in all the three cases at 21.  This value corresponds to the
hyperspherical level with $K=3$. At large distances the real part of
the $\lambda$--function goes to 5, that is the level corresponding to
$K=1$ (not reached in the figure where the maximum value of $\rho$
is only 15~fm).  The real part of the lowest rotated $\lambda$ does
not start at 5 ($K=1$) for $\rho =0$, since this $\lambda$ is Pauli
forbidden and excluded by our use of the phase equivalent
potentials. If this Pauli forbidden state had not been suppressed the
lowest $\lambda$ would start at 5, and parabolically diverge to
$-\infty$ at large distances \cite{gar99}. In the figure we also show
the imaginary part of the lowest $\lambda$'s. They behave at short and
large distances as dictated by eqs.(\ref{imal}) and
(\ref{iminf}). They start at zero, become negative according to
$-\rho^2\sin{(2\theta)}$, they cross the zero axis, and at large
distances they return to zero from the positive side.

In the inner part of the figures we compare for $\theta=0.25$ the
deepest $\lambda$ when the $1^-$ (solid lines) and $2^-$ (dashed
lines) $s$--states are at 50 keV.  We observe in the figure insets
that for the $1/2^+$ and the $3/2^+$ excited states the presence of a
low--lying $1^-$ $s$--state is making the lowest $\lambda$ slightly
deeper than when we have the $2^-$ level at 50 keV.  For the $5/2^+$
state it is the opposite, the $2^-$ level in $^{10}$Li favors a deeper
$\lambda$--function.  The reason is that $1/2^+$, but not $5/2^+$, can
be constructed when both neutron-core states simultaneously are
relative $1^-$-states and vice versa, $5/2^+$, but not $1/2^+$, can be
constructed when both neutron-core states simultaneously are relative
$2^-$-states.  For $3/2^+$ both neutrons can simultaneously be coupled
to the core in each of the $1^-$ or $2^-$ states, but the overlap is
larger for the $1^-$ than for the $2^-$-state.

\begin{figure}[ht]
\begin{center}
\epsfysize = 7cm
\epsfbox{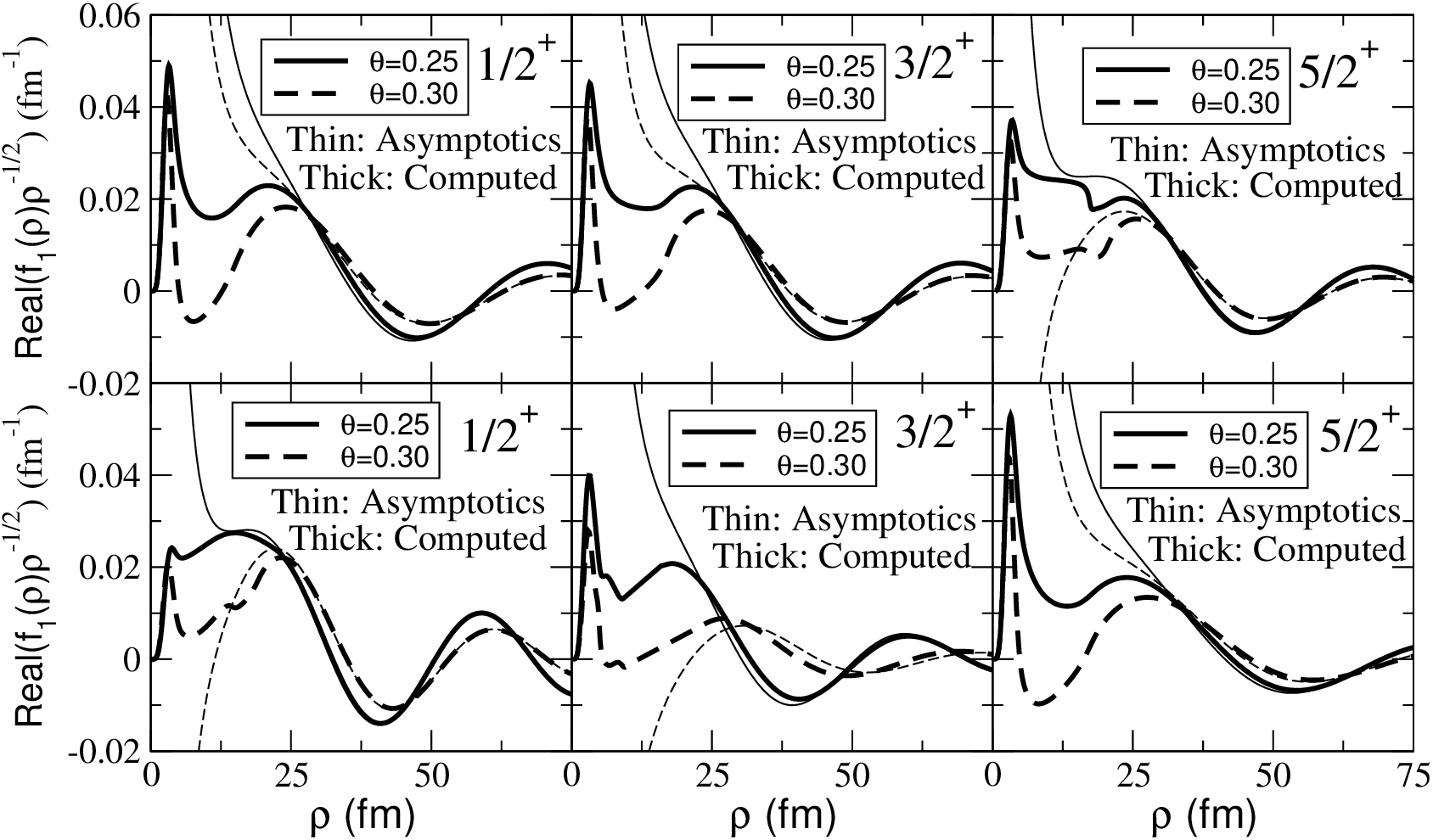}
\end{center}
\caption[]{ Real part of the rotated radial wave function associated
with the rotated lowest $\lambda$--function (see
eq.(\ref{radial})). Solid and dashed lines correspond to a rotation
angle of $\theta=0.25$, and $\theta=0.30$, respectively. Thick lines
are the computed wave functions, and the thin lines are the asymptotic
functions $H^{(1)}_{K+2}(\kappa \rho)$, see eq.(\ref{asym3b}). The
left, middle, and right parts of the figure show the wave functions
corresponding to the $1^-$ excited states in $^{11}$Li with
$J^\pi=1/2^+$, $3/2^+$ and $5/2^+$, respectively. The upper part of the
figure correspond to a $1^-$ virtual $s$--state in $^{10}$Li at 50
keV, while in the lower part a $2^-$ virtual $s$--state at 50 keV in
$^{10}$Li is assumed. In all the cases a $2^+$ $p$--resonance in
$^{10}$Li at 0.54 MeV and a $1^+$ $p$--resonance in $^{10}$Li at 0.25
MeV are assumed.}
\label{fig3}
\end{figure}

\subsection{Complex rotated radial wave functions}

\begin{figure}[ht]
\begin{center}
\epsfysize = 7cm
\epsfbox{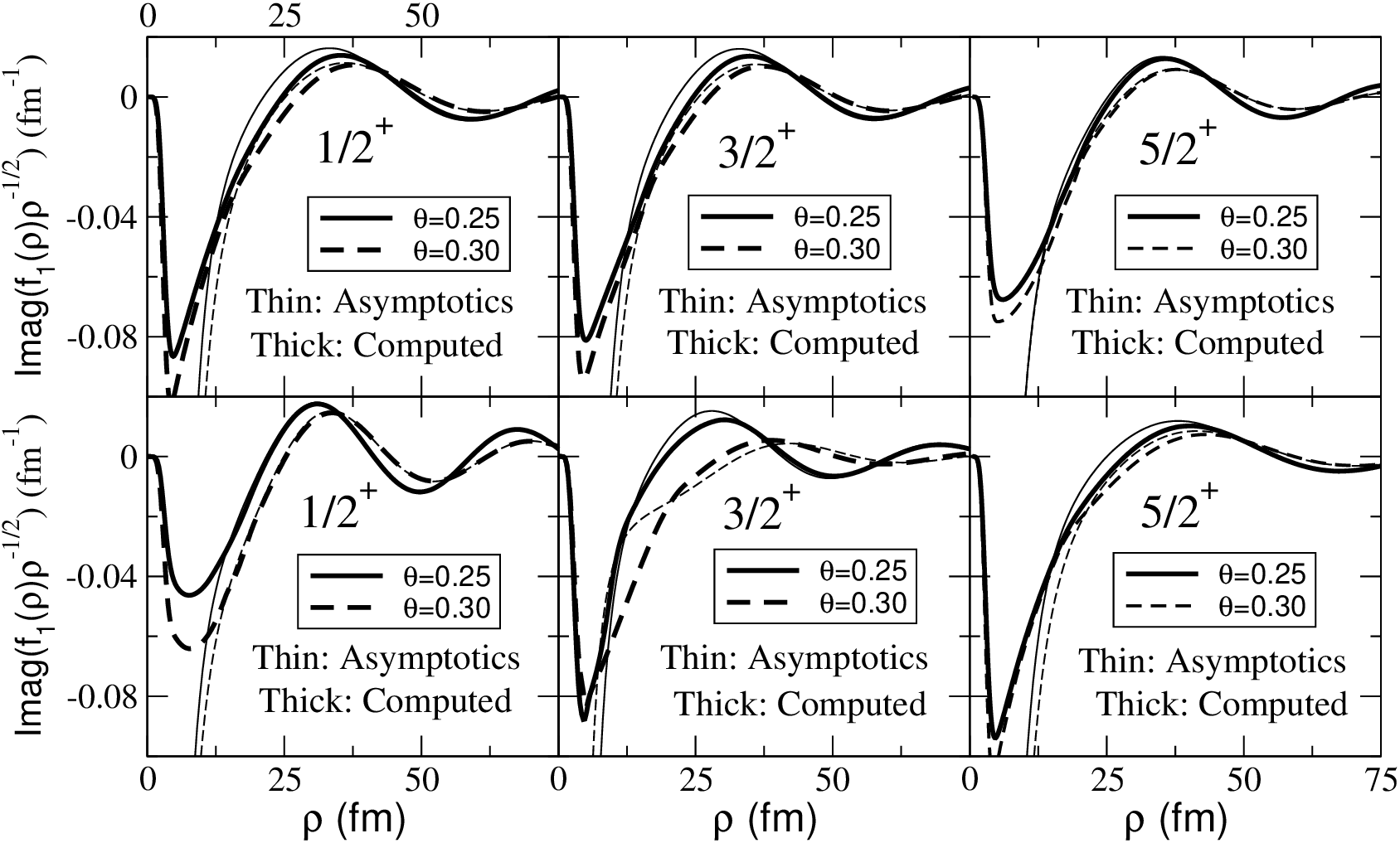}
\end{center}
\caption[]{ The same as fig.\protect\ref{fig3} but for the imaginary 
part of the rotated radial wave functions.}
\label{fig4}
\end{figure}

The effective radial potential arising from the $\lambda$--functions
in fig.\ref{fig2} are used in eq.(\ref{radial}). In figs.\ref{fig3}
and \ref{fig4} we show the real and imaginary parts of the complex
rotated radial wave function $f_1(\rho)$ associated with the deepest
$\lambda$--function.  The wave function has been divided by
$\sqrt{\rho}$. We also show the asymptotic wave functions, which
according to eq.(\ref{asym3b}) for a resonance after complex rotation
is given by $H^{(1)}_{K+2}(|\kappa|\rho e^{i(\theta-\theta_R)})$.

The asymptotics is reached at values of the hyperradius smaller than
50 fm, except for the $J^\pi=3/2^+$-state where a $2^-$ $s$--state is
at 50 keV in $^{10}$Li (middle--low part of the figure).
In this case the asymptotics is reached at
$\rho\approx 80$ fm.  From eq.(\ref{asym3br}) we know that the
exponential decrease of the rotated radial wave functions at large
distances is governed by the exponent $-|\kappa|\rho
\sin{(\theta-\theta_R)}$.  Therefore larger values of the rotation
angle $\theta$ will produce a faster decrease of the radial wave
functions as seen in both figs.\ref{fig3} and \ref{fig4}.

The complex scaled radial wave functions exhibit peaks at short
distances. In fact this is a manifestation of the three--body
resonance.  Due to the faster decrease of the wave functions for large
values of $\theta$ this peak is more pronounced for $\theta=0.30$ than
for $\theta=0.25$. The peak at short distances is especially narrow
and the three-body resonance is well--defined for the real part of the
radial wave functions in fig.\ref{fig3}. The peak structure at short
distances is also pronounced but much broader for the imaginary part
of the wave functions shown in fig.\ref{fig4}. In some cases the radial 
wave functions present strange kinks, as in the upper part of fig.3 
for $J^\pi=5/2^+$ and $\theta=0.25$ at $\rho \approx 15$ fm. This is 
produced by a sharp crossing in the two effective potentials 
(the $\lambda$--functions) included in eq.(11). Nevertheless this crossing 
does not influence the resonance energy as demonstrated by independence 
of $\theta$ for values where the crossing is smoother.

\subsection{Dipole excitations}

In table \ref{tab4} we show the resonance energies ($E_R$) and widths
($\Gamma_R$) of the $1^-$ excitations in $^{11}$Li obtained after
solving the Faddeev equations by use of the complex rotated
hyperspherical adiabatic method. The left part of the table (columns 2
and 3) corresponds to a neutron--$^9$Li interaction producing a $1^-$
virtual $s$--state in $^{10}$Li at 50 keV, while in the right part
(columns 4 and 5) a $2^-$ state is at 50 keV. The excitation energy
($E^\star$) is obtained after summation of the two--neutron separation
energy in $^{11}$Li to the resonance energy ($E^\star=E_R+0.3$ MeV).
These numbers have been obtained after scanning the complex plane for values 
of the complex energy up to 1.5 MeV, and
using a maximum value of the complex rotation angle $\theta$ of 0.35 rads.
We have then to be aware that resonances corresponding to higher energies or
with argument larger than 0.70 rads could have been missed in the
calculation.

\begin{table}
\caption{ Computed energies ($E_R$) and widths ($\Gamma_R$) for the lowest
excited $1^-$ states in $^{11}$Li for $J^\pi=1/2^+$, $J^\pi=3/2^+$, and
$J^\pi=5/2^+$. Columns 2 and 3 are obtained assuming a $1^-$ virtual 
$s$-state in $^{10}$Li at 50 keV, while in columns 4 and 5 it is a $2^-$
$s$--state the one at 50 keV. The energy of the $1^+$ $p$--resonance
in $^{10}$Li is 0.25 MeV in columns 2 and 4, and 0.45 MeV in columns 3
and 5.  }
\begin{tabular}{ccccc}
\hline
       & \multicolumn{4}{c}{$(E_R, \Gamma_R)$ (MeV)} \\ \cline{2-5}
       & \multicolumn{2}{c}{$1^-$ at 50 keV}  
       & \multicolumn{2}{c}{$2^-$ at 50 keV} \\ \hline
$E(1^+)$   & 0.25 MeV & 0.45 MeV  & 0.25 MeV & 0.45 MeV   \\ \hline
  $\frac{1}{2}^+$  &  (0.41, 0.26)   & (0.44, 0.29)  &   
                      (0.65, 0.59)   & (0.70, 0.65) \\
  $\frac{3}{2}^+$  &  (0.44, 0.27)   & (0.52, 0.39)  &   
                      (0.58, 0.35)   & (0.62, 0.55) \\
  $\frac{5}{2}^+$  &  (0.51, 0.31)   & (0.63, 0.46)  &   
                      (0.32, 0.15)   & (0.42, 0.25) \\ 
\hline
\end{tabular}
\label{tab4}
\end{table}

We start by assuming the $^{10}$Li structure shown in fig.\ref{fig1},
i.e. a $1^+$ $p$--resonance at 0.25 MeV and a $2^+$ $p$--resonance at
0.54 MeV.  These numbers are consistent with the experimental data
given in \cite{boh97}.  In columns 2 and 4 of table~\ref{tab4} we give
the computed $^{11}$Li resonance energies and widths for the $^{10}$Li
spectra in the left and right parts of fig.\ref{fig1},
respectively. From the resonance energies shown in table~\ref{tab4} we
see that only the 1/2$^+$ excited state in column 4, and to a lower
extent the 3/2$^+$ state in the same column, gives rise to an
excitation energy of around 1 MeV in agreement with the available
experimental values ( $E^\star=1.25\pm0.15$ \cite{kor96} and
$E^\star=1.02\pm0.07$ \cite{gor98}). The three excited states given in
column 2 have an excitation energy a bit lower, between 0.7 and 0.8
MeV.  Finally, the 5/2$^+$ state in column 4 has an excitation energy
of around 0.6 MeV, clearly below the experimental value.

The computed resonances in table~\ref{tab4} are similar to the lowest
ones obtained in \cite{cob98} by a different method but with realistic
interactions and in particular with the important hyperfine splitting.
In \cite{cob98} several resonances of the same spin and parity were
found in contrast to all other computations including the present one.
Large distances are responsible for these additional resonances.
Accurate treatment of distances well beyond 100~fm is required and no
other computation has achieved that. Unfortunately the complex
coordinate rotation increases the range of the rotated interaction and
thereby increase the difficulties of treating large distances. The
advantage is that the boundary condition for the resonance wave
functions are transformed into the exponentially vanishing bound state
condition.

Switching off the spin splitting interaction, it is still possible to
maintain essentially all three-body related $^{11}$Li ground state
structure and reaction properties \cite{gar02}. Then the two-body
resonances of $^{10}$Li all must be at the average positions indicated
in fig.\ref{fig1} around 0.4~MeV. The three $1^-$-resonances then
reduce to only one, that after the corresponding calculation without
the spin splitting neutron--core interaction is found to be at
$(E_R,\Gamma_R) = (0.48,0.33)$ MeV. This energy corresponds
approximately to the statistically averaged values of the 1/2$^+$,
3/2$^+$ and 5/2$^+$ resonance energies given in columns 2 and 4 of
table~\ref{tab4}. In these averages the weights of the $1/2^+$,
$3/2^+$, and $5/2^+$ energies are 1/6, 2/6, and 3/6, respectively,
according to the different number of spin projections in each
case. The average values are then $(0.47,0.29)$ MeV and $(0.46,0.29)$
MeV for the second and fourth columns in table~\ref{tab4},
respectively.

However, the properties of $^{10}$Li are inconsistent with the
parameter choice of zero spin splitting and two low-lying resonances
of opposite parity in the $^{10}$Li spectrum.  For the ground state of
$^{11}$Li only the average positions are important since the two
neutrons essentially are forced to couple to $0^+$ and therefore
simultaneously occupy both low and high lying two-body resonance
states. For the $1^-$-excitations essentially only one level in each
pair of these spin split states need to be occupied to produce the
excited states. Therefore the finite core spin and the subsequent spin
splitting is crucial for breaking the degeneracy of the $1^-$-states.
In some cases some of the splitted levels might be lowered
substantially compared to results of zero core-spin computations with
the same average two-body resonance positions.

It is important to emphasize that in \cite{gar02} we concluded that
the most likely spectrum for $^{10}$Li has a $1^+$ $p$--resonance with
energy of 0.35$\pm$0.15 MeV. Therefore a $1^+$ $p$--resonance at 0.25
MeV as chosen in the present calculations is a lower limit of the
predicted energy range. The effect of a higher energy for the $1^+$
resonance in $^{10}$Li can be seen in the columns 3 and 5 of
table~\ref{tab4}, where we have chosen an energy of 0.45 MeV for the
$1^+$ $p$--resonance. Of course this higher energy value is producing
higher energies for the three--body excited states. In some cases, as
for the $5/2^+$--states, the energy increase can reach even more than
100 keV. In this case the lowest excitation energy is 0.72 MeV.
Again, a calculation of the resonance energy suppressing the
spin--splitting interaction is reducing the resonances in columns 3
and 5 in table~\ref{tab4} to a single resonance with energy and width
of $(0.56,0.43)$ MeV that is similar to the statistically averaged
energies of $(0.56,0.41)$ and $(0.53,0.42)$ MeV obtained with the
resonance values in columns 3 and 5, respectively.  Therefore, by
simple comparison of the computed three--body excitation energies for
$^{11}$Li and the experimental value of 1 MeV, it seems that the $1^+$
$p$--resonance energy in $^{10}$Li is more likely at an energy higher
than the 0.25 MeV measured in \cite{boh97}.

The structure of the degenerate $1^-$-resonance for zero
spin-splitting corresponds to an equal probability for neutron-core
$s$ and $p$-waves. This is simply because essentially only $s$ and
$p$-waves contribute and a three-body $1^-$-excitation has to be made
of neutron-core $l_x=0,1$ and the corresponding $l_y=1,0$. The
antisymmetry of the two neutrons then requires roughly 50\% of both
the $s$ and $p$-wave neutron-core relative states. This equal division
is therefore independent of the average positions of the neutron-core
$p$-resonance and virtual $s$-state. The finite core-spin distributes
the probabilities on more components depending on the coupling
necessary to produce the total angular momentum of the three-body
resonance, see tables~\ref{tab1}, \ref{tab2} and \ref{tab3}.

The results shown in table~\ref{tab4} are independent of the rotation
angle used in the calculation. In particular, the numbers given in the
table have been obtained with rotation angles $\theta=0.25$,
$\theta=0.30$, and $\theta=0.35$.  Another point is the possible
three--body effects that are not taken into account in the
calculation. It is well known that computation of three--body ground
states by use of pure two--body interactions generally underbinds the
three--body system. This problem is solved by inclusion of an
attractive three--body potential in the radial equation (\ref{radial})
that accounts for the polarization of the particles that are beyond
that described by the effective two--body interactions.  The results
shown in table~\ref{tab4} have been obtained without use of a
three--body potential.

Including a three-body potential of range 3~fm essentially leaves the
resonance parameters unchanged. The reason is that the generalized
centrifugal barrier already provide a rather repulsive potential at
distances smaller than 3~fm. The three-body potential is then only
marginally changing the effective radial potential and has with this
range and a reasonable strength very little influence. This may
reflect that a three-body potential, at least for non-zero orbital
angular momentum states, should depend on other space variables than
the hyperradius, i.e. directions and relative size of the $\bd{x}$ and
$\bd{y}$ coordinates. Effectively this probably corresponds to a
larger range due to the resulting different asymmetric geometry
related to non-zero angular momentum. This also may result in a
potential with both attractive and repulsive regions.  Although the
effect still has to be small the resonance positions could then either
move up or down.

\subsection{Coulomb dissociation cross section}

The cross section for Coulomb excitation of electric dipole states in
the projectile nucleus is given by \cite{ber88}
\begin{equation}
\frac{d\sigma_c}{dE^\star}=\frac{N_{E1}(E^\star)}{E^\star} \sigma_{E1}(E^\star)=
\frac{N_{E1}(E^\star)}{\hbar c} \frac{16 \pi^3}{9} \frac{d B(E1)}{dE^\star}
\;,
\label{dsde}
\end{equation}
where $\sigma_{E1}(E^\star)$ is the photonuclear cross section, $B(E1)$
is the dipole strength function, and the
number of equivalent photons $N_{E1}(E^\star)$ is given by
\begin{equation}
N_{E1}(E^\star)=  
\frac{2}{\pi} Z_1^2 \alpha \left( \frac{c}{v}\right)^2
\left[
\xi K_0(\xi) K_1(\xi) - \frac{v^2 \xi^2}{2c^2} (K_1(\xi)^2 - K_0(\xi)^2)
\right] \;, 
\label{fotnum} 
\end{equation}
where $K_0$ and $K_1$ are the modified Bessel functions, $Z_1$ is the
charge of the projectile, $\alpha$ is the fine structure constant, and
\begin{equation}
\xi=\frac{E^\star R}{\hbar \gamma v}; \hspace*{4mm} 
R=R_1+R_2+\frac{\pi a}{2};
\hspace*{4mm} a=\frac{Z_1 Z_2 \alpha}{2 E_{kin}} \; ,
\end{equation}
$v$ is the velocity of the projectile,
$\gamma=(1-v^2/c^2)^{-1/2}$, $R_1$ and $R_2$ are the radii of
projectile and target, respectively, $Z_2$ is the charge of the target, 
and $a$ is half the distance of
closest approach between projectile and target. $E_{kin}$ is the
kinetic energy of the projectile.

The connection between the measured cross section $d\sigma_M/dE$ and
the true cross section $d\sigma_c/dE$ is given by
\begin{equation}
\frac{d\sigma_M}{dE}(E)=\int \frac{d\sigma_c}{dE^\prime}(E^\prime)
\varepsilon(E^\prime, E) dE^\prime \; ,
\label{conv}
\end{equation}
where $\varepsilon(E^\prime, E)$ represents the response of the
detector system. The decay energy $E$ is related to the excitation
energy $E^\star$ according to $E^\star=E+S_{2n}$, where $S_{2n}$ is
the two--neutron separation energy.

\begin{figure}[ht]
\begin{center}
\epsfysize = 7cm
\epsfbox{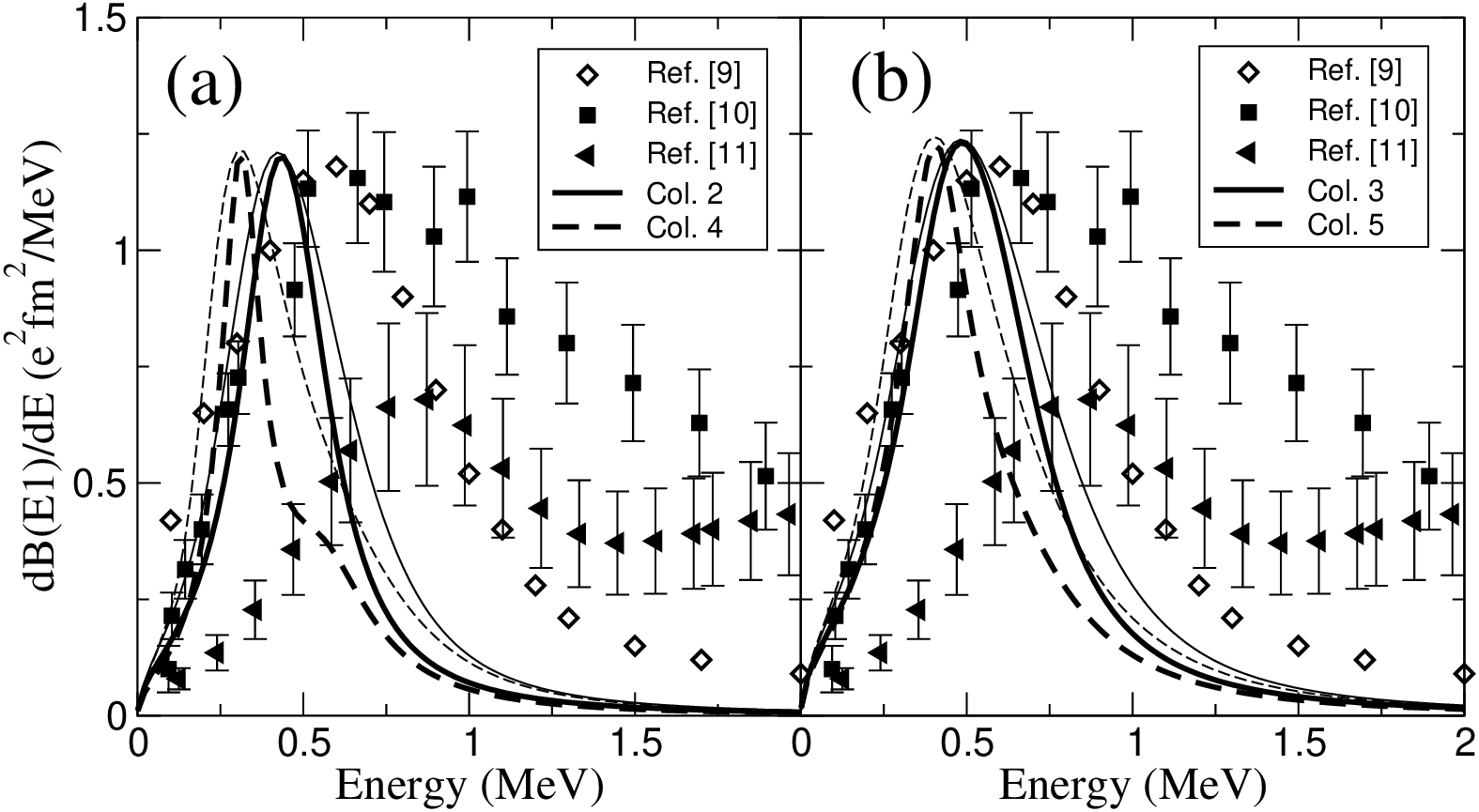}
\end{center}
\caption[]{(a) Differential B(E1)-values obtained with the three--body
resonances in columns 2 (solid line) and 4 (dashed line) in
table~\ref{tab4}. (b) The same as in the right part but for columns 3
(solid line) and 5 (dashed line). In both figures the thin lines are
the calculations after convoluting with the response of the detector
(see text).  Experimental data are from \cite{sac93}, \cite{shi95},
and \cite{zin97}, where the beam energies are 28 MeV/nucleon, 43
MeV/nucleon, and 280 MeV/nucleon, respectively.  }
\label{fig5}
\end{figure}

Assuming that the Coulomb dissociation process takes place through a
three-body resonance decay mechanism in which only the resonances are
populated in the final state, the photonuclear cross section
$\sigma_{E1}$ can be parameterized with a Breit--Wigner function given
by

\begin{equation}
\sigma_{E1}(E)=\frac{\sigma_m \Gamma(E)}{(E-E_R)^2+0.25 \Gamma(E)^2};
\hspace*{1mm}
\Gamma(E)=\Gamma_R\frac{E^{0.5}}{E_R^{0.5}}
\label{breit}
\end{equation}
where $E_R$ and $\Gamma_R$ are the energy and width of the populated
resonance.

In fig.\ref{fig5} we show the differential $B(E1)$--strength computed
as given in eq.(\ref{dsde}). The photonuclear cross section
$\sigma_{E1}$ is computed for the four sets of resonance energies in
columns 2 to 5 in table~\ref{tab4} as a weighted average of the three
Breit--Wigner functions (\ref{breit}) obtained for $J^\pi=1/2^+$
(weight=1/6), $J^\pi=3/2^+$ (weight=2/6), and $J^\pi=5/2^+$
(weight=3/6). The weight of each angular momentum $J$ is dictated by
the different number of angular momentum projections in each case. The
solid and dashed lines show the results obtained with the resonance
energies given in columns 2 and 4 of table~\ref{tab4}
(fig.\ref{fig5}a), and in columns 3 and 5 (fig.\ref{fig5}b),
respectively. The thick lines are the calculations as described above,
while the thin lines correspond to the transformation given in
eq.(\ref{conv}). The detector response function has been taken to be a
gaussian whose width fits the experimental width given in \cite{sac93}
(width=$0.402 E^{0.633}$ MeV). As seen in the figure, the convolution
(\ref{conv}) makes the curves slightly broader than the original
ones. The experimental data shown in the figure are taken from
\cite{sac93} (open diamonds) that correspond to a $^{11}$Li beam
energy of 28 MeV/nucleon, from \cite{shi95} (black squares)
corresponding to a beam energy of 43 MeV/nucleon, and from
\cite{zin97} (black triangles) where the beam energy is 280
MeV/nucleon. The experimental data given by the open diamonds
correspond to the differential $B(E1)$ strength obtained in
ref.\cite{sac93} following eqs.(\ref{dsde}) and (\ref{breit}) and
using the resonance parameters $E_R=0.7$ MeV and $\Gamma_R=0.8$ MeV
that according to ref.\cite{sac93} reproduce the experimental
decay--energy spectrum. The computed curves are all scaled to the
maximum of the experimental data.

As seen in the figure the experimental data do not agree with each
other.  At least for the sets of data from refs.\cite{sac93} and
\cite{shi95}, for which the $^{11}$Li beam energy is similar, one
would expect also a similar experimental distribution. More consistent
and accurate experimental measurements are therefore necessary. It is
also clear from the figure that the computed curves are in obvious
disagreement with the experimental data, especially when comparing
with the ones in refs.\cite{shi95} and \cite{zin97}. However, several
important points should be taken into account. First, it is
demonstrated in \cite{myo01} that in the method of complex coordinate
rotation the contributions from both two and three-body non-resonant
continuum structures are essential to obtain the full strength
function. Including only Breit-Wigner shapes around three-body
resonances omits most of the sometimes large two-body contribution.
Staying on the real energy axis as in \cite{cob98,for02} all
contributions are in principle included but of course only at the
correct energies if the correct final state continuum wave functions
and the correct reaction mechanism are used. Second, the role played
by the nuclear projectile--target interactions has not been
considered. For low beam energies this contribution is certainly
negligible while for beam energies above 200 MeV/nucleon, as the one
used in \cite{zin97}, previous calculations for $^{11}$Li on Pb show
that the nuclear contribution could amount up to 30\% of the total
two--neutron removal cross section \cite{gar00}. Finally, the
different effect produced by the possible reaction mechanisms has also
to be taken into account. Actually the different experiments are
analyzed in different ways indicating the belief in a corresponding
reaction mechanism. While in \cite{sac93} the authors assume a decay
through resonances, in \cite{shi95} the experimental data are
reproduced assuming a direct breakup mechanism. Other mechanisms, as
decay after populating two--body (neutron--core) resonances
\cite{myo01} could also be important.

\begin{figure}[ht]
\begin{center}
\epsfysize = 7cm
\epsfbox{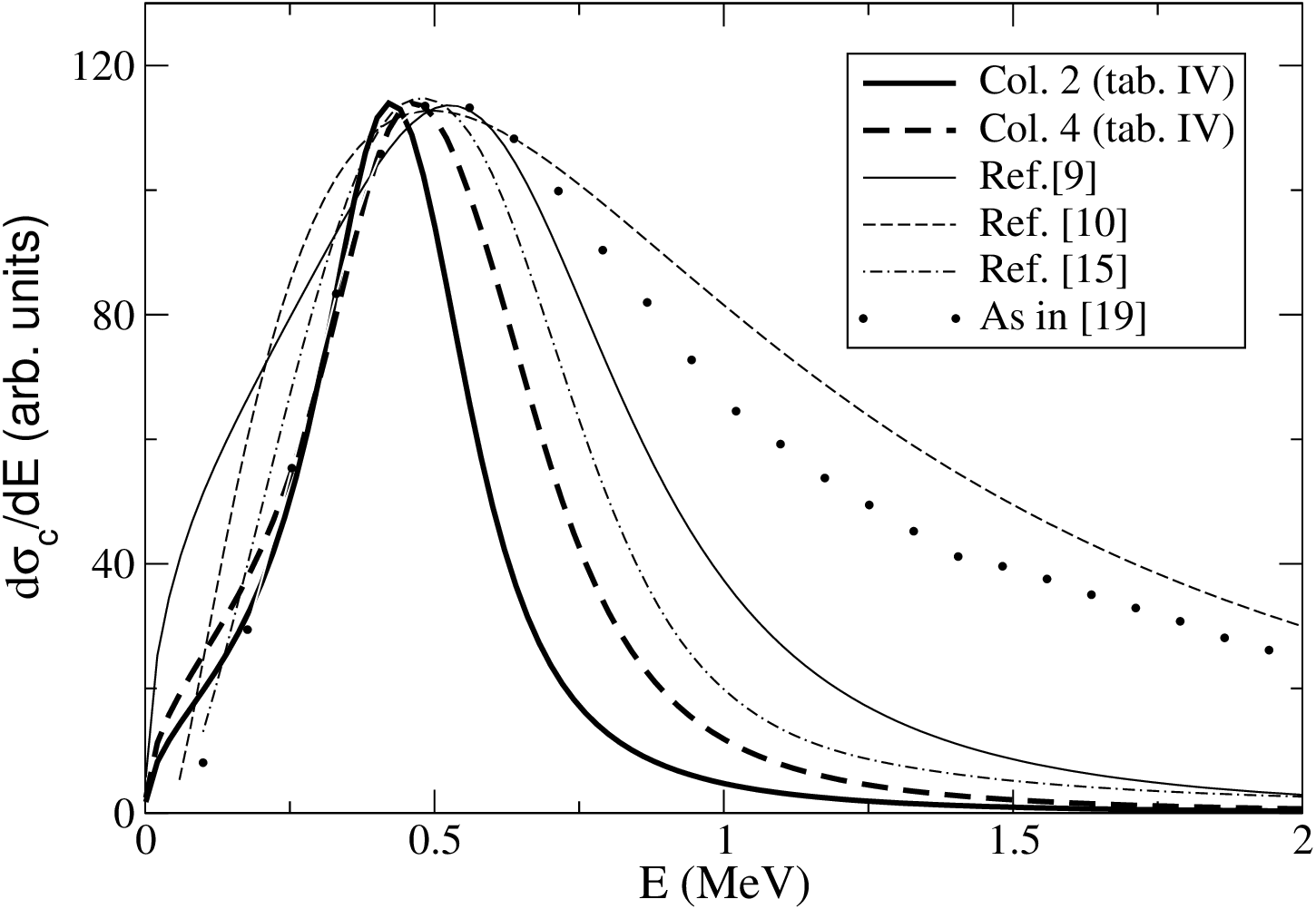}
\end{center}
\caption[]{ Differential Coulomb dissociation cross section after
fragmentation of $^{11}$Li on Pb. The thick solid and thick dashed
lines are the calculation as described in the text for the resonances
in columns 2 and 4 of table~\ref{tab4}. The thin solid line, the thin
dashed line, and the thin dot--dashed line are the calculations as
described in refs.\cite{sac93}, \cite{shi95}, and \cite{cob98},
respectively. The dotted line is the result of a calculation using
spin 0 for the $^{11}$Li projectile and no final state interactions.
This calculation is similar to the ones shown in \cite{for02}.  All
curves are scaled to the same maximum. }
\label{fig6}
\end{figure}

We can then conclude that not only experimentally, but also from the 
theoretical point of view, a careful analysis of the reaction process
is required. Reproducing the strength function with a model without the
decisive features has very little significance. To emphasize this point
we compare in fig.\ref{fig6} different calculations of the Coulomb 
dissociation cross section. Calculations assuming a decay through
three--body resonances (eqs.(\ref{dsde}) and (\ref{breit})) are shown
by the thick solid line, thick dashed line, and thin solid line, 
that correspond, respectively, to our calculations using the complex 
rotation method and using the three--body resonances shown in columns 2 
and 4 of table ~\ref{tab4}, and the calculation using the values
of $E_R$ and $\Gamma_R$ given in \cite{sac93}. The thin dashed line
is the calculation described in \cite{shi95}, that assumes a
direct breakup mechanism. This curve matches the experimental
data given in the same reference. The two remaining curves are 
calculations using a full continuum three-body wave function in the 
real energy axis. The first one (dot--dashed line) is the calculation 
described in \cite{cob98}. In this case the spin--3/2 $^{11}$Li
 ground--state wave function is obtained by solving the Faddeev 
equations by use of the hyperspeherical adiabatic method and the final 
state interactions between the three particles are included. The second
one (dotted line) is a calculation similar to the one in \cite{cob98}
but taking spin zero for the $^{11}$Li ground--state wave function
and using plane waves in the final state. These are the same conditions
as the ones used in \cite{for02}, and the computed curve is actually
similar to the ones shown in this reference. 

The pronounced differences between the computations reveals the
importance of the inclusion of all the correct ingredients: The right
reaction mechanism and the right final state interaction including not
only the final three--body continuum structures but also the
contribution from two--body continuum states.  Proper computations
could be like in \cite{myo01} with the appropriate spin splitting or as
in \cite{gar01} where the dominating decay mechanism proceeds via the
low-energy three-body continuum and the final state interaction is
incorporated by use of three-body distorted continuum wave
functions. Both calculations are rather elaborate and beyond the scope
of this paper.

\section{Summary and conclusions}
\label{sec5}

The hyperspheric adiabatic expansion has been very successful in
descriptions of ground states of three-body halo nuclei and especially
for Borromean systems. In addition a series of breakup reaction
processes are well described in the same model supplied with
appropriately specified reaction mechanisms. Complex scaling as a tool
to compute resonance energies has also been very successful and
frequently used especially in atomic and molecular physics. However,
the method has also proven efficiency in applications to nuclei,
including halo nuclei. Combining these methods of hyperspheric
adiabatic expansion and complex rotation therefore seems to be
worthwhile.

The smoking gun signal for three-body halos was first found in the
reaction cross section of $^{11}$Li and subsequently this nucleus has
been thoroughly studied. However, amazingly few realistic computations
of continuum states exist and in particular of resonances. The reasons
for this fact are that such continuum three-body computations are
difficult and especially for $^{11}$Li due to the finite spin of both
the nucleus itself and the $^{9}$Li. Furthermore the two-body
interactions are not accessible to direct measurements and
consequently not very well established.  The present level of accuracy
and sophistication demand that all observables are computed
consistently within the same model.

These theoretical problems have more or less disappeared during the
recent years. The same model has for $^{11}$Li been used for
essentially all known ground state and three-body breakup
observables. A consistent set of interaction parameters is established
in agreement with the available experimental data, but not yet applied
on continuum properties. We therefore first remove the last obstacle,
i.e. formulation of complex scaling in connection with the
hyperspheric expansion. An important ingredient is to account for the
antisymmetrization between core and valence neutrons. In our model
this is achieved by using phase equivalent two-body potentials. Thus
we first formulate complex scaling for such potentials in connection
with the present method.

After the formalism and model parameters are made available we
investigate excitations where a $1^-$ angular momentum and parity is
transfered to the ground state of $^{11}$Li. This implies three sets
of quantum numbers $J^\pi=1/2^+$, $3/2^+$ and $5/2^+$ and we search
for corresponding resonance states. The lowest adiabatic potentials
are only marginally different for the three sets of quantum numbers
with attractive pockets at relatively small values of $\rho$. The
radial wave functions are therefore very similar with peaks at small
values of $\rho$ indicating resonance properties. The eigenvalues
correspondingly reveal low-lying resonance states scattered around
the results for zero core spin (zero spin-splitting) of position above
threshold and width of (0.48, 0.33)~MeV. This corresponds to an
excitation energy position of about $0.78$~MeV, i.e. a few hundred keV
below the results extracted from measurements. The lowest of the spin
split resonances would for the most favorable interaction even lie
lower at an excitation energy of 0.74~MeV.

The resonance position is off hand low compared to the quoted
experimental value. However, other information about Coulomb
dissociation cross sections or B(E1)-strength functions is also closely
connected to the structure of the $1^-$-excitations. Three experiments
derive these quantities, but unfortunately all in mutual
disagreement. They claim and use different reaction mechanisms in the
derivations. One uses a parametrization in terms of three-body
resonances in the complex plane. This is insufficient since
significant contributions from the background continuum then are not
included. This experiment may also have acceptance problems at the
important region of small energies. Another experiment seems to assume
that the breakup is a direct process where the corresponding analysis
results in a much broader distribution. The third experiment at a
higher beam energy of 280~MeV/nucleon most probably also receives
contributions from nuclear breakup.

To get an indication we have used the computed three-body resonance
parameters and obtained the cross section.  Although this is rather
far from a correct computation we are thereby able to compare the
influence of different continuum spectra on the differential cross
section. In general our strength functions come out relatively narrow
concentrated at low energies. This is entirely consistent with the
calculated low-lying $1^-$-resonances.  Whether it also is consistent
with the measurements remains to be seen in comparison with better
experimental values and for correct computations, where the proper
reaction mechanism is used. Both these enterprises require a
substantial effort, but may well be worth doing. In this paper we have
presented our predictions of three-body $1^-$-resonances in $^{11}$Li
obtained in a model able to reproduce all ground state properties and
essentially all other three-body breakup observables.

{\bf Acknowledgements} We want to thank Karsten Riisager for
continuous discussions.

\end{document}